\documentclass[sigconf]{acmart}

\usepackage{booktabs}
\usepackage{multirow}
\usepackage{subfigure}

\usepackage{xspace}
\usepackage{url}
\usepackage{array}
\usepackage{verbatim}
\usepackage{soul, color}
\usepackage{enumitem}
\usepackage{multirow}
\usepackage{amsmath}
\usepackage{algorithm}
\usepackage{algpseudocode}

\theoremstyle{definition}
\newcommand{\eat}[1]{}
\newcommand{\ie}{\emph{i.e.,}\xspace}
\newcommand{\eg}{\emph{e.g.,}\xspace}

\newcommand{\baby}{KGCL\xspace}
\newcommand{\paratitle}[1]{\noindent\textbf{#1}}

\AtBeginDocument{%
  \providecommand\BibTeX{{%
    \normalfont B\kern-0.5em{\scshape i\kern-0.25em b}\kern-0.8em\TeX}}}

\setcopyright{acmcopyright}

\def\model{KGCL}

\begin{document}
\fancyhead{}

\title{Knowledge Graph Contrastive Learning for Recommendation}

\author{Yuhao Yang}
\affiliation{University of Hong Kong}
\email{yuhao-yang@outlook.com}

\author{Chao Huang}
\authornote{Chao Huang is the corresponding author.}
\affiliation{University of Hong Kong}
\email{chaohuang75@gmail.com}

\author{Lianghao Xia}
\affiliation{University of Hong Kong}
\email{aka\_xia@foxmail.com}

\author{Chenliang Li}
\affiliation{Wuhan University}
\email{cllee@whu.edu.cn}

\begin{abstract}
Knowledge Graphs (KGs) have been utilized as useful side information to improve recommendation quality. In those recommender systems, knowledge graph information often contains fruitful facts and inherent semantic relatedness among items. However, the success of such methods relies on the high quality knowledge graphs, and may not learn quality representations with two challenges: i) The long-tail distribution of entities results in sparse supervision signals for KG-enhanced item representation; ii) Real-world knowledge graphs are often noisy and contain topic-irrelevant connections between items and entities. Such KG sparsity and noise make the item-entity dependent relations deviate from reflecting their true characteristics, which significantly amplifies the noise effect and hinders the accurate representation of user's preference.

To fill this research gap, we design a general \underline{K}nowledge \underline{G}raph \underline{C}ontrastive \underline{L}earning framework (\model) that alleviates the information noise for knowledge graph-enhanced recommender systems. Specifically, we propose a knowledge graph augmentation schema to suppress KG noise in information aggregation, and derive more robust knowledge-aware representations for items. In addition, we exploit additional supervision signals from the KG augmentation process to guide a cross-view contrastive learning paradigm, giving a greater role to unbiased user-item interactions in gradient descent and further suppressing the noise. Extensive experiments on three public datasets demonstrate the consistent superiority of our \model\ over state-of-the-art techniques. \model\ also achieves strong performance in recommendation scenarios with sparse user-item interactions, long-tail and noisy KG entities. Our implementation codes are available at \url{https://github.com/yuh-yang/KGCL-SIGIR22}.

\end{abstract}


\copyrightyear{2022}
\acmYear{2022}
\setcopyright{acmlicensed}\acmConference[SIGIR'22]{Proceedings of the 45th International ACM SIGIR Conference on Research and Development in Information Retrieval}{July 11--15, 2022}{Madrid, Spain}
\acmBooktitle{Proceedings of the 45th International ACM SIGIR Conference on Research and Development in Information Retrieval (SIGIR'22), July 11--15, 2022, Madrid, Spain}
\acmPrice{15.00}
\acmDOI{10.1145/3477495.3532009}
\acmISBN{978-1-4503-8732-3/22/07}


\maketitle

\section{Introduction}

\begin{figure}[t]
\centering
\subfigure[Yelp2018]{
\includegraphics[width=0.31\linewidth]{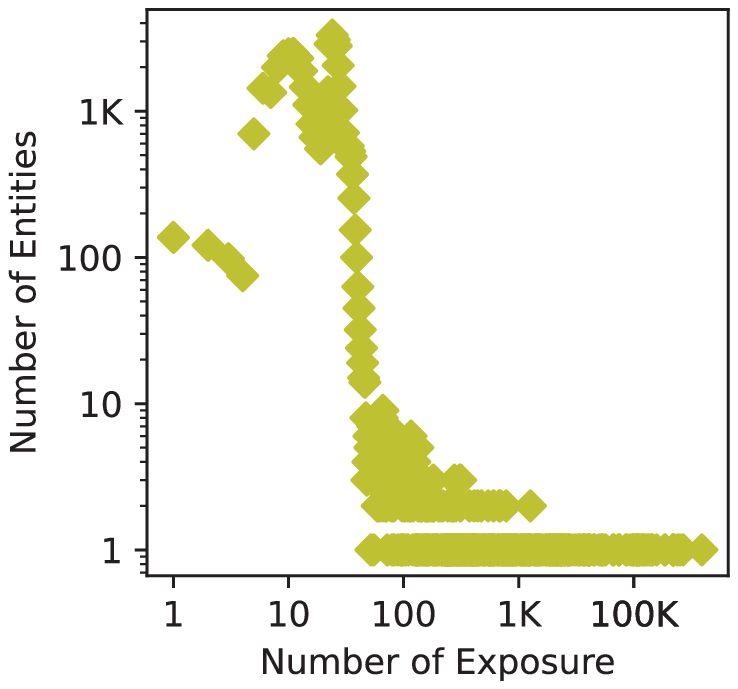}}
\subfigure[Amazon-Book]{
\includegraphics[width=0.31\linewidth]{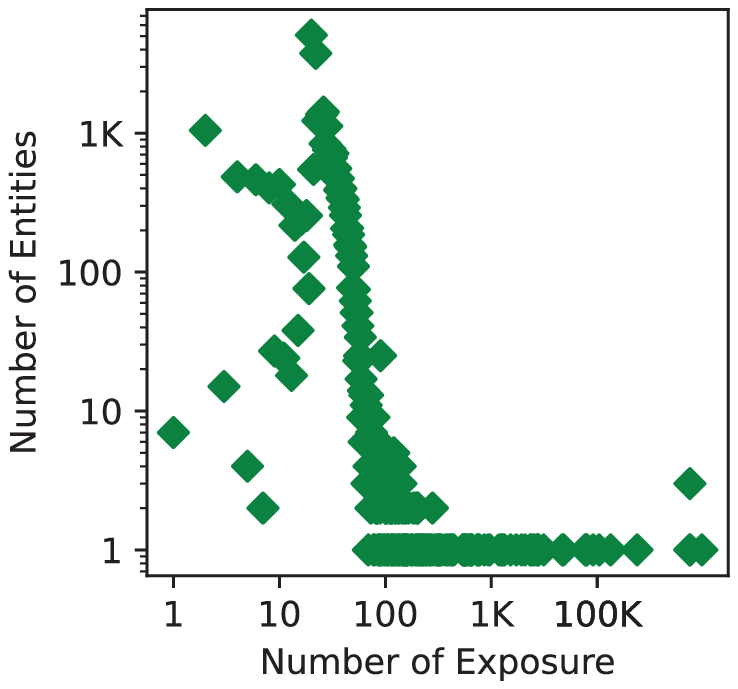}}
\subfigure[MIND]{
\includegraphics[width=0.31\linewidth]{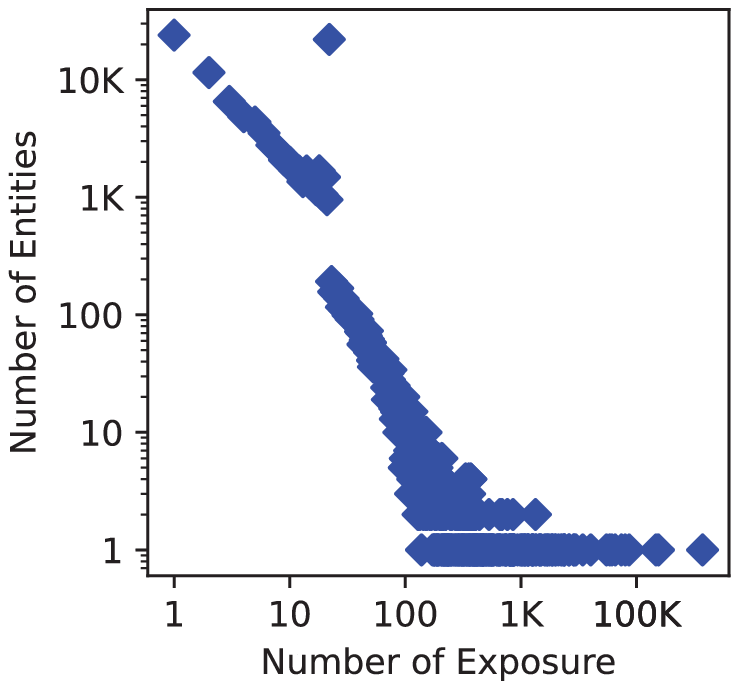}}
\vspace{-0.2in}
\caption{Long-tail entity distributions in real-world KGs.}
\label{fig:long_tail_case}
\vspace{-0.2in}
\end{figure}

Recommender systems have increasingly become an integral installation for suggesting interested items to users and alleviating information overloading in many online services, ranging from E-commerce platforms~\cite{wang2020time}, video-sharing sites~\cite{liu2019user} to online advertising~\cite{gharibshah2021user}. Among various techniques, Collaborative Filtering (CF) frameworks become effective solutions to predict users' preferences, based on the rationale that users with similar interaction behaviors may share similar interests for items~\cite{he2017wwwneural,rendle2020neural,liang2018variational}.

In recent years, the prevalent collaborative filtering paradigms have evolved from matrix factorization (MF) to neural network-based techniques for latent user and item embedding projection, such as Autoencoder-based approaches (\eg Autorec~\cite{sedhain2015autorec}), attentive CF mechanisms (\eg ACF~\cite{chen2017attentive}), as well as recently developed CF models built upon graph convolutional architectures (\eg LightGCN~\cite{he2020lightgcn}). However, even with the modeling of complex user-item interaction patterns, most CF-based recommendation methods still suffer from the data scarcity issue for users who have not yet interacted with sufficient items~\cite{togashi2021alleviating,fan2019graph,huang2021recent}. To overcome such data sparsity problem, Knowledge Graphs (KGs) serving as useful external sources have been incorporated into the recommender system to enhance the user and item representation process, by encoding additional item-wise semantic relatedness~\cite{wang2019kgat,xian2019reinforcement,social2021knowledge}.


Existing KG-enhanced methods can be roughly categorized into three groups. Particularly, some studies~\cite{cao2019unifying,zhang2016collaborative} bridge the knowledge graph learning with user-item interaction modeling, through adopting the transition-based entity embedding schemes (\eg, TransE~\cite{bordes2013translating}, TransR~\cite{lin2015learning}) to generate prior item embeddings. To improve KG-enhanced recommender systems in capturing high-order KG connectivity, some path-based models~\cite{wang2019explainable,wang2018ripplenet,zhao2017meta} aim to construct path-guided user-item connections with the incorporated KG entities. Nevertheless, most of those path-based methods involve the design of meta-paths for generating entity-dependent relationships, which requires specific domain knowledge and labor-intensive human efforts for accurate path construction. Motivated by the strength of graph neural networks, one promising recent research line lies in recursively performing information propagation among multi-hop nodes and injecting long-range relational structures, such as KGAT~\cite{wang2019kgat}, MVIN~\cite{tai2020mvin}, KHGT~\cite{xia2021knowledge} and KGIN~\cite{wang2021learning}.

Despite their effectiveness in some scenarios, we argue that the effectiveness of existing KG-aware recommendation methods largely relies on the high quality input knowledge graphs and are vulnerable to noise perturbation. However, in practical scenarios, knowledge graphs are often sparse and noisy by exhibiting long-tail entity distribution and containing topic-irrelevant connections between items and entities~\cite{pujara2017sparsity,wang2018label}.

\begin{figure}
	\centering
	\includegraphics[width=0.95\columnwidth]{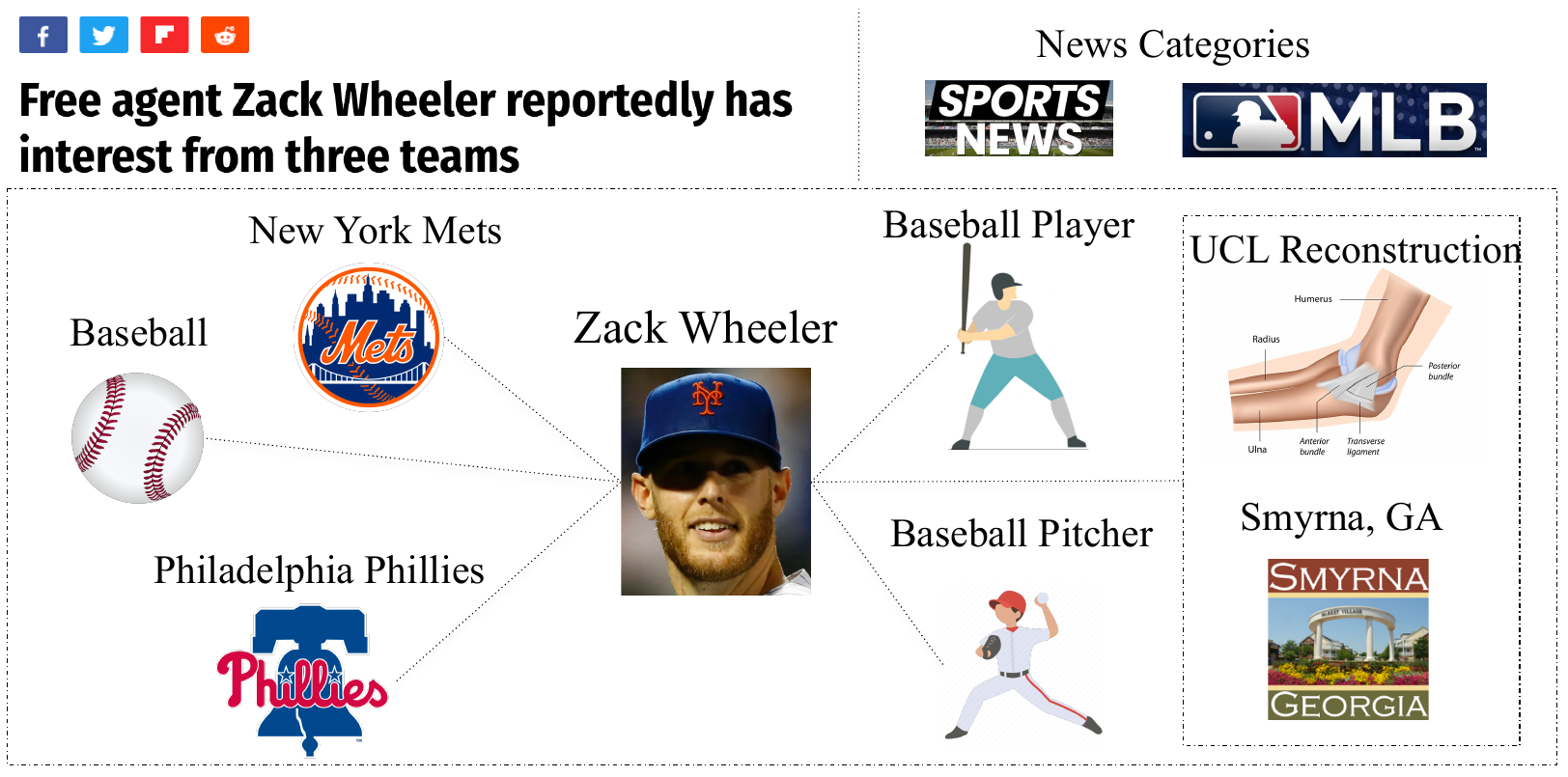}
	\vspace{-0.2in}
	\caption{Illustrated example of news topic-irrelevant entities extracted from the knowledge graph of MIND dataset.}
	\vspace{-0.2in}
	\label{figure:intro_noise}
\end{figure}

We report distributions of KG entities collected from three real-world datasets in Figure \ref{fig:long_tail_case} to illustrate the long-tail issue in KGs. In this figure, the y-axis represents the number of entities corresponding to the exposure count in the x-axis. Obviously, across datasets from different platforms, \ie food, book and news, the majority of KG entities exhibit long-tail phenomenon. Since it requires enough triplets $(h,r,t)$ linked to an entity to accurately model semantic transitions in the KG by employing Trans algorithms \cite{bordes2013translating, lin2015learning}, which brings challenge for accurately capturing item-wise relatedness.
Additionally, topic-irrelevant entity connections are ubiquitous in KGs. We present a motivating example for news recommendation shown in Figure~\ref{figure:intro_noise}, the key entity \textit{Zack Wheeler} extracted from the news item is a famous baseball pitcher for the Philadelphia Phillies in the organization of Major League Baseball (MLB). However, we can notice that \textit{Zack Wheeler} is linked with two item semantic-irrelevant ``noisy'' entities, \ie \textit{Smyrna, GA} and \textit{UCL Reconstruction}. While \textit{Zack Wheeler} was born in \textit{Smyrna, GA} and he underwent a surgery for Ulnar Collateral Ligament (UCL) reconstruction before, these two entities are less relevant to the main topic of this news with the focus on the recent sports news.

We unify the aforementioned problems as the KG noise issue. Such data noise issue will impair the quality of item representation from two perspectives: i) From the local view, directly aggregating information from those low-quality entities will bring noise in preserving key semantics of items from their neighboring entities. ii) From the global view, the information aggregation over the knowledge graph is easily towards over-smoothed, since the overwhelming information can be propagated to the target node through some popular entities (\eg location names). For example, other persons who are also born in \textit{Smyrna, GA} can be connected to the professional baseball pitcher--\textit{Zack Wheeler}. Hence, it is a necessity to endow the knowledge graph-enhanced CF paradigm with the capability of effectively connection denoising, so as to distill the true underlying preference of target users with representations invariant to noise disturbance.

\noindent \textbf{Contribution}. In light of the aforementioned limitations and challenges, we propose a general \underline{K}nowledge \underline{G}raph \underline{C}ontrastive \underline{L}earning framework (\model) for recommendation. Specifically, to handle the relation heterogeneity in knowledge graph, we first propose a relation-aware knowledge aggregation mechanism to capture the entity- and relation-dependent contextual signals for the first-stage item representation.
Then, we develop a cross-view contrastive learning schema which bridges the knowledge graph denoising with the user-item interaction modeling, such that the external item semantic relatedness can be leveraged to guide the data augmentation with cross-view self-supervised signals. The designed cross-view contrastive learning schema suppresses KG noise via performing KG contrastive learning, and exploits external signals from the process to measure the bias of item representations affected by KG noise. The signals serve as the guidance to the user-item graph contrastive learning view, keeping useful graph structures and involve less noise.

\model\ takes inspirations from the knowledge graph learning and self-supervised data augmentation, to incorporate the knowledge graph context to guide the model in refining user/item representations with new knowledge-aware contrastive objectives. In our framework, our joint contrastive learning network learns to drop irrelevant KG triplets and items based on the knowledge graph structural consistency, for robust user preference learning. Due to the model-agnostic property of our \model, it can be plugged into various graph neural recommendation models. In \model, the knowledge graph-guided contrastive learning model and graph neural CF architecture are jointly optimized in an end-to-end manner.

In summary, our contributions are highlighted as follows:

\begin{itemize}[leftmargin=*]

\item This work introduces the idea of integrating the knowledge graph learning with user-item interaction modeling under a joint self-supervised learning paradigm, to improve the robustness and alleviate the data noise and sparsity issues for recommendation.

\item We present a general \model, a knowledge graph-guided topological denoising framework, offering cross-view self-discrimination supervision signals with knowledge-aware contrastive objective. We also provide theoretical analysis to justify the benefits brought by the integrative learning objective.

\item We conduct diverse experiments on three public datasets and the proposed \model\ consistently outperforms various state-of-the-art recommendation methods across different settings. Further ablation analysis justifies the rationality of our key components.


\end{itemize}

\vspace{-0.1in}
\section{Preliminaries}

This section introduces key notations used throughout the paper and formalize our studied task. We consider a typical recommendation scenario with a user set $\mathcal{U}$ and an item set $\mathcal{I}$. Individual user and item is denoted as $u$ ($u\in \mathcal{U}$) and $i$ ($i\in \mathcal{I}$), respectively. We define the user-item interaction matrix $\mathcal{Y} \in \mathcal{|U|\times |\mathcal{I}|}$ to represent the consumption behaviors of users over different items. In matrix $\mathcal{Y}$, the element $y_{u,i}=1$ given that user $u$ has adopted item $i$ before (\eg click, review or purchase) and $y_{u,i}=0$, otherwise. \\\vspace{-0.12in}

\noindent \textbf{User-Item Interaction Graph}. Based on the matrix $\mathcal{Y}$, we first construct the user-item interaction graph $\mathcal{G}_u = \{\mathcal{V}, \mathcal{E}\}$, where the node set $\mathcal{V} = \mathcal{U} \cup \mathcal{I}$ and edge $(u,i)$ is generated in $\mathcal{G}$ if $y_{u,i}=1$. \\\vspace{-0.12in}

\noindent \textbf{Knowledge Graph}. We let $\mathcal{G}_k = \{(h,r,t)\}$ represent the knowledge graph which organizes external item attributes with different types of entities and corresponding relationships. Specifically, each entity-relation-entity triplet $(h,r,t)$ characterizes the semantic relatedness between the head and tail entity $h$ and $t$ with the relation $r$, such as the triplet for movie recommendation (Titanic, Directed by, James Cameron), and venue recommendation (McDonald's, Located in, Chicago). Such information incorporates fruitful facts and connections among items as side information to improve the modeling of user preference for recommendation.

Having constructed user interaction behaviors and item knowledge, we seek to leverage the item knowledge information to assist the user interest learning. However, real-world knowledge graphs are often noisy and involve item-irrelevant entities as we described before. In such cases, not all entities and relations are useful for learning appropriate item characteristics. In most existing knowledge-aware recommender systems, messages aggregated from ``noisy''entiries and relations may heavily impact the quality of item representation, which notably limits the effectiveness of KG-enhanced user preference modeling. To tackle this challenge, this work exploits the potential of knowledge graph topology denoising to distill informative guidance for user and item representations. \\\vspace{-0.12in}

\noindent \textbf{Task Formulation}. We formally describe our task as follows: \textbf{Input}: user-item interaction data $\mathcal{G}_u = \{\mathcal{V}, \mathcal{E}\}$ and item knowledge graph data $\mathcal{G}_k = \{(h,r,t)\}$. \textbf{Output}: the learned function $\mathcal{F} = (u,v|\mathcal{G}_u, \mathcal{G}_k, \Theta)$ that forecasts the items user $u$ ($u\in \mathcal{U}$) would like to interact with, where $\Theta$ denotes the model parameters.

\vspace{-0.1in}
\section{METHODOLOGY}
\label{sec:method}
\begin{figure*}
    \centering
    \includegraphics[width=1.0\linewidth]{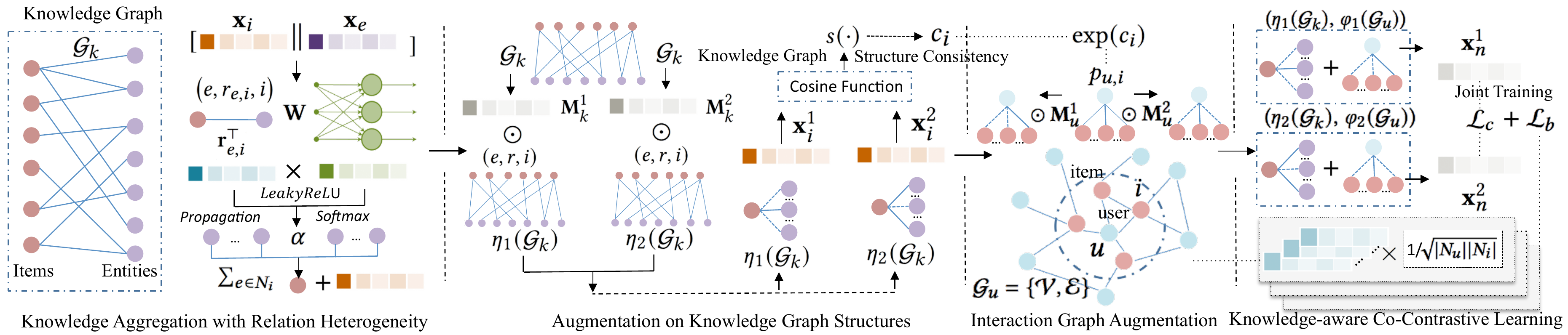}
    \vspace{-0.25in}
    \caption{The overall architecture of our proposed \model. Knowledge-aware co-contrastive learning with augmentation functions on both knowledge graph $\eta(\cdot)$ and user-item interaction graph $\varphi(\cdot)$. Our contrastive objective $\mathcal{L}_{c}$ is jointly optimized with main embedding space shared by the knowledge graph aggregation and graph-based CF encoder.}
    \vspace{-0.15in}
    \label{fig:framework}
\end{figure*}

We present the overall architecture of \model\ in Figure~\ref{fig:framework}. Technical details are discussed in following sub-sections.


\subsection{Relation-aware Knowledge Aggregation}
\label{rgat}

\subsubsection{\rm \textbf{Knowledge Aggregation with Relation Heterogeneity}} Inspired by the graph attention mechanisms in~\cite{velivckovic2017graph,wang2019kgat,xia2021knowledge}, we first design a relation-aware knowledge embedding layer to be reflective of relation heterogeneity over knowledge graph connection structures. To address the limitation of manually design of path generation on knowledge graphs, our \model\ projects entity- and relation-dependent context into specific representations with parameterized attention matrix. Towards that, we built our message aggregation mechanism between the item and its connected entities in $\mathcal{G}_k$, for generating knowledge-aware item embeddings based on the heterogeneous attentive aggregator shown as follows:
\begin{align}
	\textbf{x}_{i}&=\textbf{x}_{i}+\sum_{e \in \mathcal{N}_{i}} \alpha\left(e, r_{e,i},i\right) \textbf{x}_{e} \nonumber\\ 
	\alpha\left(e, r_{e,i},i \right)&=\frac{\exp \left({ LeakyReLU }\left(\mathbf{r}_{e,i}^{\top}\mathbf{W}\left[\textbf{x}_{e} \| \textbf{x}_{i}\right]\right)\right)}{\sum_{e \in N_{i}} \exp \left({ LeakyReLU }\left(\mathbf{r}_{e,i}^{\top}\mathbf{W}\left[\textbf{x}_{e} \| \textbf{x}_{i}\right]\right)\right)}
	\label{eq:kge}
\end{align}
\noindent where $N_{i}$ is the neighboring entities of item $i$ based on different types of relations $r(e,i)$ in knowledge graph $\textbf{G}_{k}$. Here, the embedding of item and entity is denoted as $\textbf{x}_i \in \mathbb{R}^{d}$ and $\textbf{x}_e \in \mathbb{R}^{d}$, respectively. $\alpha\left(e, r_{e,i},i \right)$ represents the estimated entity- and relation-specific attentive relevance during the knowledge aggregation process. In particular, $\alpha\left(e, r_{e,i},i \right)$ encodes the distinct semantics of relationships between item $i$ and entity $e$. $\mathbf{W}\in \mathbb{R}^{d\times 2d}$ represents the parametric weight matrix customized to the input item and entity representations. \emph{LeakyReLU} activation function is adopted for non-linear transformation.

\subsubsection{\rm \textbf{Semantic Representation Enhancement}} Additionally, to further enhance the multi-relational semantic representation space for entity-item dependencies, we perform the alternative training between our relation-aware knowledge aggregator and TransE~\cite{bordes2013translating}. The general idea of this translation-based knowledge graph embedding is to make the summation of head and relation embedding $\textbf{x}_h$ and $\textbf{x}_r$ as close as the tail representation $\textbf{x}_t$. Here, we define $f_d(\cdot)$ to represent the $L_1$ norm-based similarity measurement function between embedding vectors, \ie $f_d = \left \| \textbf{x}_h + \textbf{x}_r - \textbf{x}_t \right \|$. Formally, the translation-based optimized loss $\mathcal{L}_{TE}$ is shown below:
\begin{align}
\mathcal{L}_{TE} = \sum_{(h,r,t,t^\prime)\in \mathcal{G}_k} -\ln \sigma \left( f_d (\textbf{x}_h, \textbf{x}_r, \textbf{x}_{t^\prime}) - f_d (\textbf{x}_h, \textbf{x}_r, \textbf{x}_t) \right)
\end{align}
\noindent The negative sample $t^\prime$ is generated by randomly replacing the tail $t$ for the observed triplets $(h,r,t)$ from the knowledge graph $\mathcal{G}_k$.

\subsection{Knowledge Graph Augmentation}


\subsubsection{\rm \textbf{Augmentation on Knowledge Graph Structures}}
Motivated by the recent success of data augmentation techniques with contrastive learning in CV/NLP tasks, \eg image analysis~\cite{verma2021towards} and machine translation~\cite{ruiter2019self}, we propose to bridge the knowledge graph embedding and the contrastive learning paradigm with auxiliary self-supervised signals. At the core of contrastive learning is to maximize the mutual information between augmented views and regularize the embedding learning with contrastive objectives.

In our \model\ framework, we propose to generate different views of knowledge graph structures for contrastive learning through the entity-wise self-discrimination. In particular, we adopt stochastic data augmentation scheme over the input knowledge graph to generate two correlated data views. Then, the knowledge graph structural consistency of individual item is derived to be reflective of item-wise invariance to knowledge noise perturbation. Towards this end, we hence devise the data augmentation operator $\eta(\cdot)$ on the knowledge graph structure with two stochastic selections $\eta_1(\mathcal{G}_{k})$ and $\eta_2(\mathcal{G}_{k})$, which can be formally presented as follows:
\begin{align}
\eta_1(\mathcal{G}_{k}) = ((e, r, i) \odot \textbf{M}^1_k),~~~ \eta_2(\mathcal{G}_{k}) = ((e, r, i) \odot \textbf{M}^2_k)
\end{align}
\noindent where $(e, r, i) \in \mathcal{G}_{k}$ represents the knowledge triplet between items and their dependent entities. Here, we define masking vectors $\textbf{M}^1_k$, $\textbf{M}^2_k \in \{0,1\}$ as the binary indicators with the probability $p_k$, to denote whether the specific knowledge triplet is selected or not during the sampling. By doing so, we can generate knowledge subgraph with different augmented structural views. The objective of our knowledge graph augmentation scheme is to identify items which are less sensitive to structure variation, and more tolerant to the connections with noisy entities. Such identified items are less ambiguous in terms of their characteristics and are more helpful to capture the preference of the correlated users.

\subsubsection{\rm \textbf{Agreement between Augmented Structural Views}}
After performing the augmentation on knowledge graph structures, we obtain two knowledge graph dependency views with operators $\eta_1(\cdot)$ and $\eta_2(\cdot)$. Inspired by the investigation of graph consistency in~\cite{zhuang2018dual,jin2021hierarchical}, to explore the agreement property of each item based on the augmented views, we define knowledge graph structure consistency $c_i$ of item $i$ with the agreement between the representations encoded from different views as follows:
\begin{equation}
c_{i} = s \left(f_{k}\left(\mathbf{x}_{i}, \eta_1(\mathcal{G}_{k}) \right), f_{k}\left(\mathbf{x}_{i}, \eta_2(\mathcal{G}_{k}) \right)\right)
\end{equation}
\noindent Here, $f_{k}$ represents the relation-aware knowledge aggregation scheme (defined in Eq~\ref{eq:kge}), to generate item embeddings $\textbf{x}_i^1$ and $\textbf{x}_i^2$ corresponding to different augmented structure views $\eta_1(\mathcal{G}_{k})$ and $\eta_2(\mathcal{G}_{k})$. $s(\cdot)$ denotes the cosine function to estimate the similarity between $\textbf{x}_i^1$ and $\textbf{x}_i^2$. Based on the above definitions, we can notice that if an item achieves a higher structure consistency score $c_i$, it is less sensitive to the topological information changes. Therefore, if item $i$ is affected more by KG noise than item $i'$, it is more likely that $c_{i} < c_{i'}$. Such derived knowledge structure consistency property of each item can be adopted as the guidance, so as to against both the knowledge graph dependency and the user-item interaction noise with auxiliary self-supervised signals.

\subsection{Knowledge-Guided Contrastive Learning}
We integrate our knowledge graph augmentation schema with the graph contrastive learning paradigm, with the aim of improving the representation ability of graph-based collaborative filtering in terms of model accuracy and robustness. To effectively transfer useful item external knowledge in learning of user preference, we design two contrastive representation spaces for user-item interactions. In such contrastive learning framework, the denoising item knowledge can be leveraged to guide the user and item representation and alleviate the sparsity of supervision signals.

\subsubsection{\rm \textbf{Interaction Graph Augmentation Mechanism}}
While the recent proposed self-supervised recommendation model SGL~\cite{wu2021self} performs data augmentation on user-item interaction graph, the purely randomly dropout operations limits its effectiveness in keeping useful interactions for contrastive learning.

To mitigate this limitation, we leverage the estimated knowledge graph structure consistency of items to guide the data augmentation over the user-item interaction graph $\mathcal{G}_u = \{\mathcal{V}, \mathcal{E}\}$. The rationale behind our knowledge-guided graph contrastive learning is to identify interactions which are more useful to characterize user preference with less bias information. To be specific, the items with higher KG structure consistency scores will involve less noise and contribute more to the modeling of user's real interests. In accordance with our knowledge-guided augmentation, we incorporate the derived item-specific KG structure consistency $c_i$ into our operator on the user-item interaction graph with the following formulas: 
\begin{align}
\label{eq:ks}
\begin{split}
	w_{u,i} &= \exp (c_{i});~p_{u,i}^\prime = \max \left(\frac{w_{u,i}-w^{min}}{w^{max}-w^{min}},
p_\tau\right) \\
p_{u,i} &= p_a \cdot \mu_{p^\prime} \cdot p^\prime_{u,i}
\end{split}
\end{align}
\noindent where $p_{u,i}$ represents the estimated probability to dropout the interaction edge between user $u$ and item $i$. $w_{u,i}$ represents the influence degree of item $i$ over user $u$, which is proportional to the corresponding structure consistency score of $c_i$. We further perform the min-max normalization on $w_{u,i}$ with the truncation probability $p_\tau$, to alleviate the low value effect. After that, the intermediate variable $p^\prime_{u,i}$ is obtained and integrated with the mean value $\mu_{p^\prime}$ to derive the value of dropout probability $p_{u,i}$. Here, $p_a$ controls the strength of mean-based influence. With the probability $p_{u,i}$, we further generate two masking vectors $\textbf{M}^1_u$, $\textbf{M}^2_u \in \{0,1\}$ based on the Bernoulli distribution~\cite{marshall1985family}. After that, $\textbf{M}^1_u$, $\textbf{M}^2_u \in \{0,1\}$ are applied to the user-item interaction graph $\mathcal{G}_u$ as follows:
\begin{align}
\label{eq:ui_mask}
\varphi(\mathcal{G}_u) = (\mathcal{V}, \textbf{M}^1_u \odot \mathcal{E}),~~~\varphi(\mathcal{G}_u) = (\mathcal{V}, \textbf{M}^2_u \odot \mathcal{E})
\end{align}
\noindent where $\varphi(\cdot)$ denotes our graph augmentation operator which drops out the user-item interaction in the edge set $\mathcal{E}$ of graph $\mathcal{G}_u$ according to the inferred probability $p_{u,i}$.

\subsubsection{\rm \textbf{Knowledge-aware Co-Contrastive Learning}}
Different from most existing contrastive learning models (\eg SGL~\cite{wu2021self}, GraphCL~\cite{you2020graph}) which directly performs transformation on structure-level augmentations (\eg node dropping out or edge edge perturbation), we incorporate item knowledge semantics into a co-contrastive learning architecture. Our \model\ aims to improve the model robustness with augmented self-supervision signals, by disturbing the graph structures from the views of both item semantics and user behavioral patterns. In our co-contrastive learning paradigm, we integrate our designed graph augmentation operators $\eta(\mathcal{G}_{k})$ and $\varphi(\mathcal{G}_u)$ to create two contrastive views, which enables view-specific encoders collaboratively supervise with each other.

Specifically, given the obtained augmented knowledge subgraphs through $\eta_1(\mathcal{G}_{k})$ and $\eta_2(\mathcal{G}_{k})$, we further separately corrupt the user-item interaction graph guided by the derived knowledge structure consistency of items ($c_i$, $i\in \mathcal{I}$) based on $\varphi_1(\mathcal{G}_u)$ and $\varphi_2(\mathcal{G}_u)$. After that, we can create two knowledge-guided corrupted graphs among users, items and entities. Then, we encode the representations of users and items by utilizing the graph-based collaborative filtering framework and the relation-aware knowledge aggregation mechanism. Due to the effectiveness and lightweight architecture of LightGCN~\cite{he2020lightgcn}, we adopt its message propagation strategy to encode the collaborative effects from user-item interactions as below:
\begin{align}
\textbf{x}_u^{(l+1)} = \sum_{i\in N_u} \frac{\textbf{x}_i^{(l)}}{\sqrt{|N_u||N_i|}};~~~~\textbf{x}_i^{(l+1)} = \sum_{u\in N_i} \frac{\textbf{x}_u^{(l)}}{\sqrt{|N_i||N_u|}}
\end{align}
\noindent where $\textbf{x}_u^{(l)}$ and $\textbf{x}_i^{(l)}$ represents the encoded representations of user $u$ and item $i$ under the $l$-th graph propagation layer. $N_u$ and $N_i$ denotes the set of user $u$'s interacted items and item $i$'s connected users, respectively. In the graph-structured CF architecture, the high-order collaborative signals can be captured via stacking multiple graph propagation layers. In this encoding pipeline, our designed heterogeneous attentive aggregator (defined in Eq~\ref{eq:kge}) is employed to generate input item feature vector with the preservation of knowledge graph semantics. Such item embeddings are fed into the graph-based CF for representation refinement.

After that, \model\ employs the generated two knowledge-aware graph views, \ie ($\eta_1(\mathcal{G}_{k})$, $\varphi_1(\mathcal{G}_u)$) and ($\eta_2(\mathcal{G}_{k})$, $\varphi_2(\mathcal{G}_u)$), to collaboratively supervise each other. Particularly, \model\ performs contrastive learning on view-specific user/item representations ($\textbf{x}_u^1$, $\textbf{x}_i^1$) and ($\textbf{x}_u^2$, $\textbf{x}_i^2$). For each user or item node, the positive pairs are generated from the two view-specific embeddings of user $u$ or item $i$ based on the self-discrimination ability of node. Negative pairs are the representations of different nodes in both graph views. The contrastive objective $\mathcal{L}_c$ in our \model\ is defined based on the InfoNCE~\cite{chen2020simple} loss as follows:
\begin{align}
\mathcal{L}_c = \sum_{n \in \mathcal{V}} - \log \frac{\exp(s(\textbf{x}_n^1, \textbf{x}_n^2)/\tau)}{\sum_{n'\in \mathcal{V},n'\neq n} \exp(s(\textbf{x}_n^1, \textbf{x}_{n'}^2)/\tau)}
\end{align}
\noindent where $\tau$ is the temperature parameter. We adopt the cosine function $s(\cdot)$ to estimate the similarity of positive pairs and negative pairs. By minimizing the contrastive objective loss $\mathcal{L}_c$, we can achieve the agreement between positive pairs as compared to negative ones.

\noindent \textbf{Joint Training}. In the learning process of \model, we design a joint embedding space which is shared by the main recommendation task and the auxiliary self-supervised signals. In particular, we further couple the original Bayesian personalized ranking (BPR) recommendation loss with the aforementioned contrastive loss. Firstly, we formally present the employed BPR loss $\mathcal{L}_{b}$ as follows:
\begin{equation}
    \mathcal{L}_{b} = \sum_{u \in \mathcal{U}} \sum_{i \in \mathcal{N}_u} \sum_{i' \notin \mathcal{N}_u} - \log \sigma (\hat{y}_{u,i} - \hat{y}_{u,i'})
\end{equation}
\noindent where $\mathcal{N}_u$ represents the observed interactions of user $u$. We sample the negative instance from the non-interacted items $i'$ ($i' \notin \mathcal{N}_u$) of user $u$. $\hat{y}$ is the estimated interaction probability between user $u$ and item $i$, which is derived with the dot-product as: $\textbf{x}_u^\top \textbf{x}_i$.
Given above definitions, the integrative optimization loss of our \model\ is: 
\begin{equation}
    \mathcal{L} = \mathcal{L}_{b}+\lambda_1\mathcal{L}_{c}+\lambda_2\| \Theta \|_2^2,
\end{equation}
\noindent where $\lambda_1$ and $\lambda_2$ denote parameters to determine the strength of self-supervised signals and regularization for the joint loss function. $\Theta$ represents the learnable model parameters.

\subsection{Model Analysis of \model}

\subsubsection{\rm \textbf{Theoretical Discussion of \model}}

\begin{figure}
    \centering
    \includegraphics[width=\linewidth]{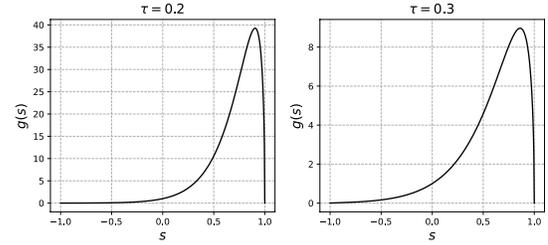}
    \vspace{-0.28in}
    \caption{Distribution of gradient function $g(s)$ under $\tau=0.2$ and $\tau=0.3$. $s$ is the similarity score between positive and negative instances. Hard negatives significant impact $g(s)$.}
    \vspace{-0.25in}
    \label{fig:gx}
\end{figure}

In our \model, we leverage the extracted knowledge graph semantics to guide the graph contrastive learning across different structural views, to improve the discrimination ability of hard negatives. Specifically, following the work in \cite{khosla2020supervised,wu2021self}, given the estimated similarity between one node instance (\ie user $u$ or item $i$) and its negative sample $v$, the obtained contrastive gradient $g(s)$ can be formally presented:
\begin{equation}
g(s)=\sqrt{1-s^{2}} \exp \left(\frac{s}{\tau}\right)
\end{equation}
\noindent We show the distribution of $g(s)$ in Figure~\ref{fig:gx} given that $\tau = 0.2$ and $\tau = 0.3$. As shown in this figure, the hard negatives with high similarity scores (\eg $0.7 \leq s \leq 0.9$), results in the value of gradient $g(s)$ close to 40. In such case, hard negatives will have a much larger influence on the gradient learning as compared to easy negatives.

In our knowledge-aware contrastive learning paradigm, the derived knowledge graph structural consistency of items is incorporated into the augmentation on user-item interaction graph, to guide the edge dropout operation. By doing so, the discrimination ability of hard negatives in \model\ can be improved from the following aspects: i) items connected with noisy entities are differentiated better with the dropout of noisy triplets \eg $\left(e, r_{e,i},i \right)$; ii) users interacted with ambiguous items can be modeled with lower similarities. Motivated by the research work in~\cite{khosla2020supervised,wu2021self}, we provide the knowledge-aware contrastive gradient analysis. Particularly, we first define the set of false hard negatives as $U_{f}$ with the definition:
\begin{equation}
    \|\theta_f(U_f, v)-s_0 \| < t_0
\end{equation}
\noindent where the largest value of $g(s)$ is obtained with the maximum point ($s_0$, $g(s_0)$). $\|\theta_f(U_f, v)-s_0 \| = \frac{\sum_{u_f \in U_{f}}|\theta_f(u_f, v) - x_0|}{|U_f|}$ measures the average similarity between node instances in $U_f$ and the corresponding positive sample $v$. Here $\theta_f$ denotes the similarity estimated with bias involved by the noisy entity-dependent information in knowledge graph. Without the knowledge graph denoising for data augmentations, the similarity between the positive and negative samples is likely to impact the model optimization with large gradient (caused by false hard negatives $U_f$) according to the distribution curve of $g(s)$:
\begin{equation}
    \|\theta_r(U_f, v)-s_0 \| \gg t_0
\end{equation}
\noindent Here, $\theta_r(\cdot)$ represents the actual estimated similarity with false hard negatives. In our \model, the effect of false hard negatives can be alleviated with our contrastive augmentation functions $\eta(\cdot)$ and $\varphi(\cdot)$. The similarity derived by \model\ is denoted as $\theta_a$. Based on above discussion, we respectively update the similarity derivation for false hard negatives $U_f$ and true hard negatives $U_t$ as follows:
\begin{align}
\label{eq:false_negative}
    \|\theta_r(U_f, v)-s_0 \| \geq \|\theta_a(U_f, v)-s_0 \| \gg \|\theta_f(U_f, v)-s_0 \| \nonumber\\
    \|\theta_f(U_t, v)-s_0 \| \gg \|\theta_a(U_t, v)-s_0 \| \geq \|\theta_r(U_t, v)-s_0 \|
\end{align}
\noindent After enhancing the discrimination ability over hard negatives, we can improve the robustness of knowledge-aware recommender systems with the accurate and helpful gradients for model learning.

\subsubsection{\rm \textbf{Model Time Complexity Analysis}}
We analyze the time complexity from three key components of our \model\ framework. (1) For the knowledge aggregation module, $O(|\mathcal{E}_k| \times d)$ calculations are required to calculate weights $\alpha$ and conduct information aggregation, where $|\mathcal{E}_k|$ denotes the number of relations in the knowledge graph $\mathcal{G}_k$. This module takes additional $O(B_t\times d)$ time for knowledge graph embedding with TransE, where $B_t$ representing the number of training triplets in a batch. (2) Our designed knowledge graph augmentation only takes $O(|\mathcal{E}_k| + |\mathcal{V}|\times d)$ time to derive the KG structure consistency and perturbation. (3) The graph-based collaborative filtering takes $O(|\mathcal{E}|\times d)$ time for user-item interaction modeling. The time complexity to calculate InfoNCE loss $O(B_{id} \times (|\mathcal{U}| + |\mathcal{I}|) \times d)$, where $B_{id}$ is the number of unique users and items within a batch. Based on the above analysis, our \model\ achieves comparable time complexity when competing with state-of-the-art knowledge-aware recommendation models~\cite{wang2019kgcn, wang2019kgat}.

\vspace{-0.1in}
\section{Experiments}



Extensive experiments are performed to evaluate the performance of our \model\ by answering the following research questions:
\begin{itemize}[leftmargin=*]

\item \textbf{RQ1}: How does our \model\ perform when competing with different types of recommendation methods?

\item \textbf{RQ2}: How do different key modules in our \model\ framework contribute to the overall performance?

\item \textbf{RQ3}: How effective is the proposed \model\ model in alleviating data sparsity and noise issues for recommendation? 

\item \textbf{RQ4}: How is the model interpretation ability of our \model?

\end{itemize}

\begin{table}
	\centering
	\footnotesize
	\caption{Statistics of experimented datasets.}
	\vspace{-0.15in}
	\label{tab:stats}
	\begin{tabular}{c|c|c|c}
		\toprule
		Stats. & Yelp2018 & Amazon-Book & MIND \\
		\hline
		\# Users & $45, 919$ & $70, 679$ & $300, 000$ \\
		\# Items & $45,538$ & $24, 915$ & $48, 957$ \\
		\# Interactions & $1, 183, 610$ & $846, 434$ & $2, 545, 327$ \\
		Density Degree & $5.7\times 10^{-4}$ & $4.8 \times 10^{-4}$ & $1.7 \times 10^{-4}$
		\\
		\midrule
		& \multicolumn{3}{c}{Knowledge Graph} \\
		\midrule
		\# Relations & $42$ & $39$ & $90$ \\
		\# Entities & $47, 472$ & $29, 714$ & $106, 500$ \\
		\# Triples & $869, 603$ & $686, 516$ & $746, 270$ \\
		\hline
	\end{tabular}
	\vspace{-0.15in}
\end{table}

\vspace{-0.1in}
\subsection{Experimental Settings}

\subsubsection{\rm \textbf{Datasets}}
We perform experiments on three public datasets collected from different real-life platforms: \emph{Yelp2018} for business venue recommendation, \emph{Amazon-Book} for product recommendation, and \emph{MIND} for news recommendation. Table~\ref{tab:stats} presents the statistical information of our experimented datasets with different interaction sparsity degrees and knowledge graph characteristics. We follow the similar settings in~\cite{wang2019kgat} to construct knowledge graphs for Yelp2018 and Amazon-Book datasets by mapping items into Freebase entities~\cite{zhao2019kb4rec}. In our experiments, we only collect entities within two hops since few of the baselines consider modeling multi-hop relations in the KG, and such relations are usually noisy and semantically biased. In our knowledge graphs, various types of entities (\eg venue category/location, book authors/publisher) are adopted to generate entity-dependent relations. For the news MIND dataset, we follow the data pre-processing strategy in~\cite{tian2021joint} to construct the knowledge graph based on \textit{spacy-entity-linker} tool\footnote{\url{https://github.com/egerber/spaCy-entity-linker}} and Wikidata\footnote{\url{https://query.wikidata.org/}}. Evaluated datasets are available in our released model implementations with the link in the abstract section.




\subsubsection{\rm \textbf{Evaluation Protocols}}
For fair comparison, we employ the all-ranking strategy to be consistent with the settings in~\cite{wang2019kgat,wang2021learning}. Specifically, for each target user, we regard all his/her non-interacted items as negative samples to infer the preference of this user. For the performance evaluation, two representative metrics: Recall@N and NDCG@N are used to evaluate the accuracy of top-$N$ recommended items~\cite{wang2019kgat,xia2021graph}. Average evaluation results across all users in the test set are reported with $N=20$ by default.

\subsubsection{\rm \textbf{Baselines for Comparison}}
We compare \model\ with various lines of recommender systems for performance evaluation.

\noindent \textbf{Conventional Collaborative Filtering Method}.
\begin{itemize}[leftmargin=*]
\item \paratitle{BPR}~\cite{rendle2012bpr}. It is a representative recommendation approach to rank item candidates with a pairwise ranking loss.
\end{itemize}


\noindent \textbf{MLP-based Neural Collaborative Filtering Framework}.\vspace{-0.05in}
\begin{itemize}[leftmargin=*]
\item \paratitle{NCF}~\cite{he2017neural}. It utilizes the multiple-layer perceptron to endow the CF architecture with the non-linear feature interaction.
\end{itemize}

\noindent \textbf{Graph Neural Networks for Collaborative Filtering}.\vspace{-0.05in}
\begin{itemize}[leftmargin=*]
    \item \paratitle{GC-MC}~\cite{berg2017graph}. It is built on the graph auto-encoder architecture to capture the interaction patterns between user and item based on the links in the bipartite graph.
    \item \paratitle{LightGCN}~\cite{he2020lightgcn}. This is a state-of-the-art GCN-based recommendation method which simplifies the convolution operations during the message passing among users and items.
\end{itemize}

\noindent \textbf{Self-Supervised Learning Recommender System}.\vspace{-0.05in}
\begin{itemize}[leftmargin=*]
\item \paratitle{SGL}~\cite{wu2021self}. This method offers state-of-the-art performance by enhancing the graph-based CF framework with augmented structure-based self-supervised signals. 
\end{itemize}

\noindent \textbf{Embedding-based Knowledge-aware Recommendation}.\vspace{-0.05in}
\begin{itemize}[leftmargin=*]
\item \paratitle{CKE}~\cite{zhang2016collaborative}. This method adopts TransR to encode the items’ semantic information and further incorporate it into the denoising auto-encoders for item representation with knowledge base.
\end{itemize}

\noindent \textbf{Path-based Knowledge-aware Recommendation}.\vspace{-0.05in}
\begin{itemize}[leftmargin=*]
\item \paratitle{RippleNet}~\cite{wang2018ripplenet}. It propagates user preference over the knowledge graph along with the constructed paths rooted at this user. It is a memory-like neural model to improve user representations.
\end{itemize}

\noindent \textbf{KG-enhanced Recommendation with GNNs}.\vspace{-0.05in}
\begin{itemize}[leftmargin=*]
	\item \paratitle{KGCN} \cite{wang2019kgcn}. It aims to encode high-order dependent context with respect to the semantic information in KG. At the core of KGCN is to incorporate neighborhood information bias into aggregating message for entity representation.
	\item \paratitle{KGAT} \cite{wang2019kgat}. This model designs an attentive message passing scheme over the knowledge-aware collaborative graph for embedding fusion. The relevance of neighboring nodes are differentiated during the propagation process. 
    \item \paratitle{KGIN} \cite{wang2021learning}. It is a recently proposed KG-enhanced recommendation model to identify latent intention of users, and further performs the relational path-aware aggregation for both user-intent-item and KG triplets.
	\item \paratitle{CKAN} \cite{wang2020ckan}. It introduces a heterogeneous propagation mechanism to determine the importance of knowledge-aware neighbors, so as to integrate the collaborative filtering representation space with the knowledge graph embedding.
	\item \paratitle{MVIN} \cite{tai2020mvin}. It is a multi-view item embedding network based on graph neural architecture. Information from both user and entity side is considered to learn feature embeddings of items.
\end{itemize}

\paratitle{Parameter Settings.}
Our proposed \model\ is implemented with PyTorch. Most of compared baselines are evaluated based on the unified recommendation library \textit{RecBole} \cite{zhao2021recbole}. In particular, we fix the embedding dimensionality as 64 for all methods, and conduct the model optimization with the learning rate of $1e^{-3}$ and batch size of 2048. For knowledge-aware recommendation models, the number of context hops and memory size is set as 2 and 8, respectively. In our \model, we search the temperature parameter $\tau$ and contrastive loss balance parameter $\lambda_1$ in the range of \{0.1,...,0.5,...,1.0\} with an increment of 0.1. Additionally, truncation probability $p_\tau$ and $p_a$ are searched among the range of $\{0.6,0.7,0.8,0.9 \}$.



\begin{table}[t]
	\centering
	\caption{Performance comparison of all methods on Yelp, Amazon and MIND. The superscript $\ast$ indicates the improvement is statistically significant where $p$-value $<0.01$ level.}
	\vspace{-0.15in}
	\label{tab:results}
	\resizebox{\linewidth}{!}{
	\begin{tabular}{c|cc|cc|cc}
		\hline
		\multirow{2}{*}{Model} & \multicolumn{2}{c|}{Yelp2018} & \multicolumn{2}{c|}{Amazon-book} & \multicolumn{2}{c}{MIND} \\
		~ & Recall & NDCG & Recall & NDCG & Recall & NDCG \\
		\hline
		\hline
		BPR & 5.55\%$^\ast$ & 0.0375$^\ast$ & 12.44\%$^\ast$ & 0.0658$^\ast$ & 9.38\%$^\ast$ & 0.0469$^\ast$ \\ 
		NCF & 5.35\%$^\ast$ & 0.0346$^\ast$ & 10.33\%$^\ast$ & 0.0532$^\ast$ & 8.93\%$^\ast$ & 0.0436$^\ast$ \\ 
		GC-MC & 6.88\%$^\ast$ & 0.0453$^\ast$ & 10.64\%$^\ast$ & 0.0534$^\ast$ & 9.84\%$^\ast$ & 0.0491$^\ast$ \\
		LightGCN & 6.82\%$^\ast$ & 0.0443$^\ast$ & 13.98\%$^\ast$ & 0.0736$^\ast$ & 10.33\%$^\ast$ & 0.0520$^\ast$ \\
		SGL & 7.19\%$^\ast$ & 0.0475$^\ast$ & 14.45\%$^\ast$ & 0.0766$^\ast$ & 10.32\%$^\ast$ & 0.0539$^\ast$ \\
		CKE & 6.86\%$^\ast$ & 0.0431$^\ast$ & 13.75\%$^\ast$ & 0.0685$^\ast$ & 9.01\%$^\ast$ & 0.0382$^\ast$ \\
		RippleNet & 4.22\%$^\ast$ & 0.0251$^\ast$ & 10.58\%$^\ast$ & 0.0549$^\ast$ & 8.58\%$^\ast$ & 0.0407$^\ast$ \\
		KGCN & 5.32\%$^\ast$ & 0.0338$^\ast$ & 11.11\%$^\ast$ & 0.0569$^\ast$ & 8.87\%$^\ast$ & 0.0431$^\ast$ \\
		KGAT & 6.75\%$^\ast$ & 0.0432$^\ast$ & 13.90\%$^\ast$ & 0.0739$^\ast$ & 9.07\%$^\ast$ & 0.0442$^\ast$ \\
		KGIN & 7.12\%$^\ast$ & 0.0462$^\ast$ & 14.36\%$^\ast$ & 0.0748$^\ast$ & 10.44\%$^\ast$ & 0.0527$^\ast$ \\
		CKAN & 6.89\%$^\ast$ & 0.0441$^\ast$ &13.80\%$^\ast$ & 0.0726$^\ast$ & 9.91\%$^\ast$ & 0.0499$^\ast$\\
		MVIN & 6.91\%$^\ast$ & 0.0441$^\ast$ & 13.98\%$^\ast$ & 0.0742$^\ast$ & 9.62\%$^\ast$ & 0.0487$^\ast$ \\
		\hline
		\hline
		\textbf{\baby} & \textbf{7.56\%} & \textbf{0.0493} & ~\textbf{14.96\%} & \textbf{0.0793} & \textbf{10.73\%} & \textbf{0.0551} \\
		\hline
	\end{tabular}
	}
	\vspace{-0.1in}
\end{table}

\subsection{Performance Comparison with SOTA (RQ1)}

We report the overall performance evaluation of all methods in Table~\ref{tab:results}. From the results, we summarize the following observation:

\begin{itemize}[leftmargin=*]

\item \model\ consistently performs better than other baselines in all cases, which verifies the effectiveness of integrating knowledge graph embedding into the contrastive learning paradigm. The diversity of evaluation datasets varying by sparsity degrees, knowledge graph characteristics, and recommendation scenarios. The superior results justify the generality and flexibility of our \model\ framework. Overall, the improvements obtained by \model\ can be attributed to two aspects: i) Benefiting from our knowledge graph contrastive learning, \model\ can denoise entity-dependent relationships and capture accurate item-wise semantics. ii) \model\ is able to guide the interaction data augmentation schema for self-supervised information with the distilled item knowledge.

\item We can observe that most of knowledge-aware recommender systems achieve better performance as compared to BPR and NCF. This confirms the helpfulness of incorporating knowledge graph information to tackle the sparsity issue in collaborative filtering. Among various knowledge-aware methods, KGIN performs the best by enhancing user representation with the exploration of the latent intention, based on the intent-aware relational paths for embedding propagation. The performance gap between our \model\ and other knowledge-aware models (\eg KGAT, CKAN, MVIN), suggests that the noisy knowledge graph misleads the learning of item-item semantic relatedness.


\item The relatively superior performance achieved by SGL indicates the rationality of generating self-supervised signals from unlabeled user behaviors, to improve the robustness of recommendation. Different from the self-supervised recommendation model SGL, our \model\ creates contrastive self-supervision signals with knowledge-guided augmentation schema, which effectively incorporates the KG-based item semantic relatedness to alleviate the interaction sparsity issue in a robust and explicit manner.

\end{itemize}

\begin{table}[t]
	\centering
	\small
	\caption{Impact study of knowledge-aware graph augmentation schema with model variants of \model.}
	\vspace{-0.15in}
	\label{tab:ablation}
	\begin{tabular}{l|cc|cc}
		\hline
		\multirow{2}{*}{Model} & \multicolumn{2}{c|}{Amazon-Book} & \multicolumn{2}{c}{MIND} \\
		~ & Recall & NDCG & Recall & NDCG \\
		\hline
		\baby & \textbf{14.96\%} & \textbf{0.0793} & \textbf{10.73\%} & \textbf{0.0551} \\
		\baby~ w/o KGA & 14.85\% & 0.0788 & 10.57\% & 0.0546 \\
		\baby~ w/o KGC  & 14.68\% & 0.0771 & 10.35\% & 0.0537 \\
		\hline
	\end{tabular}
	\vspace{-0.1in}
\end{table}

\begin{table}[t]
    \caption{Impact of $\tau$ and $\lambda_1$ on Amazon-Book dataset.}
    \vspace{-0.15in}
    \label{tab:hp}
    \centering
    \scriptsize
    \resizebox{0.85\linewidth}{!}{
    \begin{tabular}{c|c|c|c|c|c}
    \hline
        Metric & \multicolumn{5}{c}{Recall@20}  \\
        \hline
        $\lambda_1$, $\tau$ & 0.1 & 0.2 & 0.3 & 0.4 & 0.5  \\
        \hline
        \hline
        $10^{-1}$ & 12.93\% & \textbf{14.96\%} & 14.46\% & 13.94\% & 13.17\% \\
        $10^{-2}$ & 13.74\% & 13.68\% & 13.08\% & 12.39\% & 11.55\% \\
        $10^{-3}$ & 12.77\% & 11.94\% & 11.27\% & 10.62\% & 9.97\% \\
        \hline
    \end{tabular}}
\end{table}

\subsection{Ablation Study of \model\ Framework (RQ2)}
\label{subsec:ablation}

\paratitle{Impact of Knowledge-aware Graph Augmentation Schema}.
We investigate the effect of our knowledge-aware graph augmentation schema from the views of both knowledge graph and user-item interaction behaviors. accordingly, we design two model variants:

\begin{itemize}[leftmargin=*]
\item i) ``w/o KGA'': the variant of \model\ without the knowledge-guided augmentation scheme on user-item interaction graph. Instead, the contrastive views of interaction graph are constructed with randomly edge sampling, for mutual information estimation.
\item ii) ``w/o KGC'': we remove the knowledge graph contrastive learning component from \model, and directly forward the item representations encoded from our relation-aware knowledge aggregator into the graph-based CF framework for contrastive learning.
\end{itemize}

From results in Table~\ref{tab:ablation}, it is clear that the performance superiority of our \model\ framework can be achieved in all cases. This fact indicates that the proposed knowledge-guided contrastive learning over the interaction graph, and the knowledge graph contrastive learning are both effective for making better recommendations. \\\vspace{-0.12in}



\paratitle{Hyper-parameter Sensitivity.}
We further present the evaluation results of our hyperparameters of $\lambda_1$ and $\tau$ for controlling the strength of contrastive regularization and hard negative sampling, respectively. In particular, $\lambda_1$ and $\tau$ are searched from the range of ($10^{-1}$, $10^{-2}$, $10^{-3}$) and (0.1, 0.2, 0.3, 0.4, 0.5), respectively. We can observe that the best performance can be achieved by $\lambda_1=0.1$ and $\tau=0.2$, which indicates that larger value of $\tau$ may limits the discrimination ability between different negative instances. In addition, small value of $\lambda_1$ corresponds to the less influence of contrastive optimization loss on the main embedding space.

\begin{figure}[t]
	\subfigure[Amazon-Book]{
		\includegraphics[width=0.46\columnwidth]{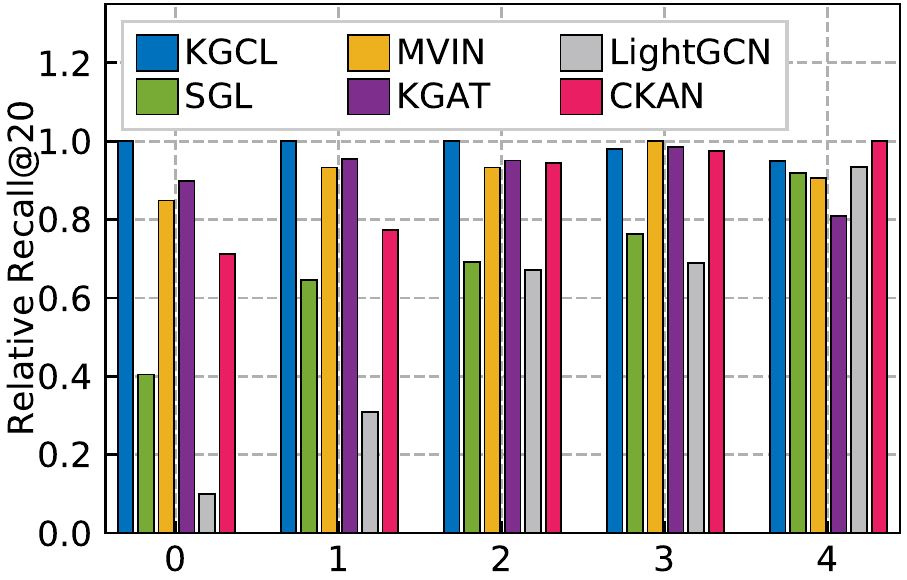}
	}
	\subfigure[Yelp2018]{
	    \includegraphics[width=0.46\columnwidth]{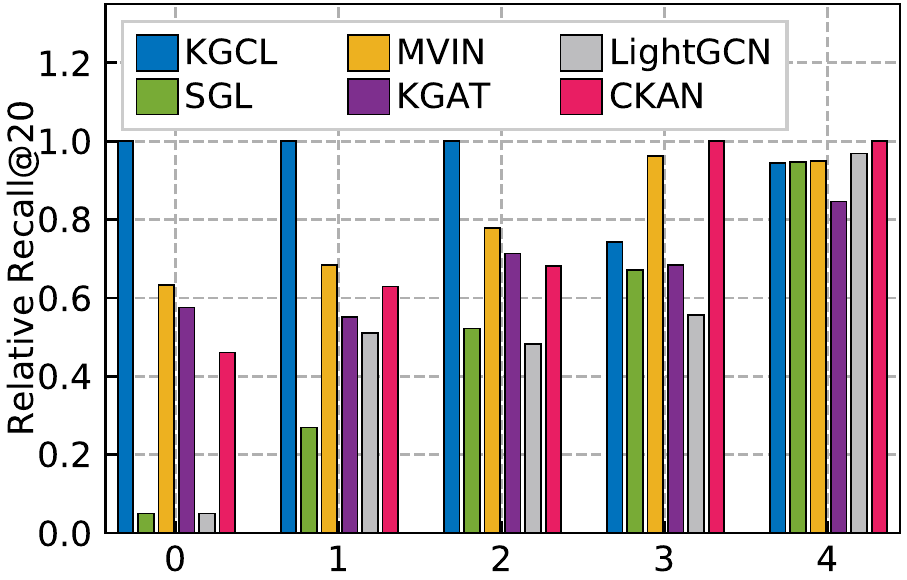}
	}
	\vspace{-0.2in}
	\caption{Performance with different interaction density degrees of items between \model\ and baselines. Recall values are normalized to range [0, 1] for better presentation.}
	\label{figure:longtail}
	\vspace{-0.2in}
\end{figure}

\begin{figure}[t]
	\subfigure[Recall]{
		\includegraphics[width=0.4\columnwidth]{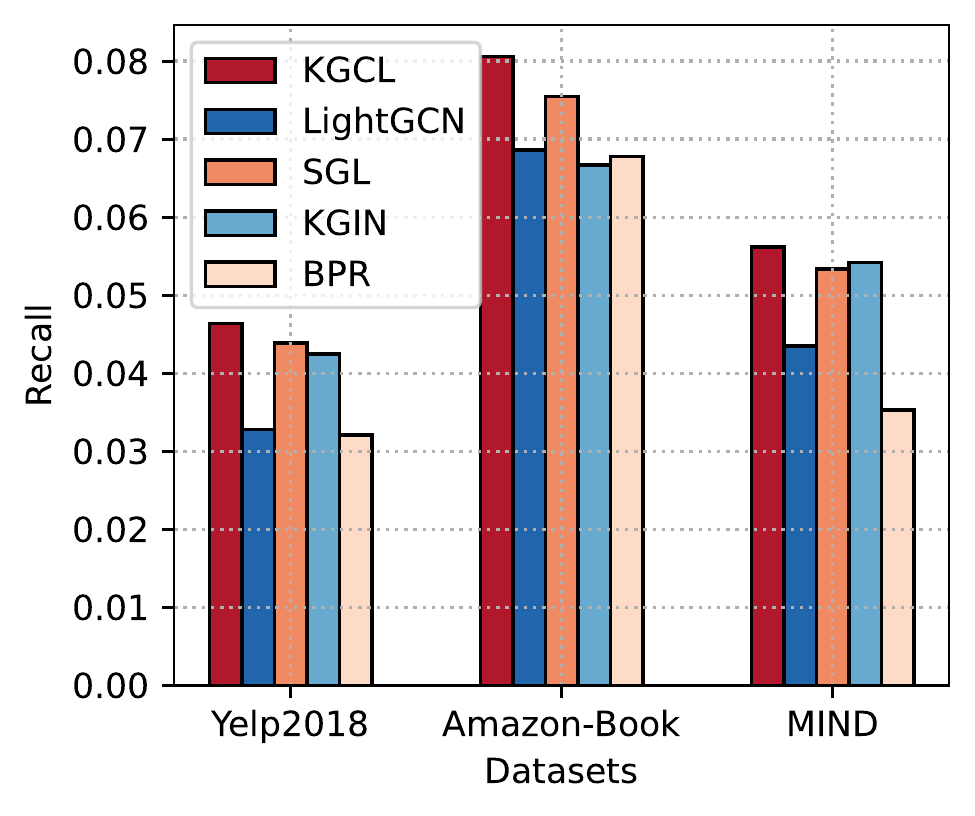}
	}
	\subfigure[NDCG]{
	    \includegraphics[width=0.4\columnwidth]{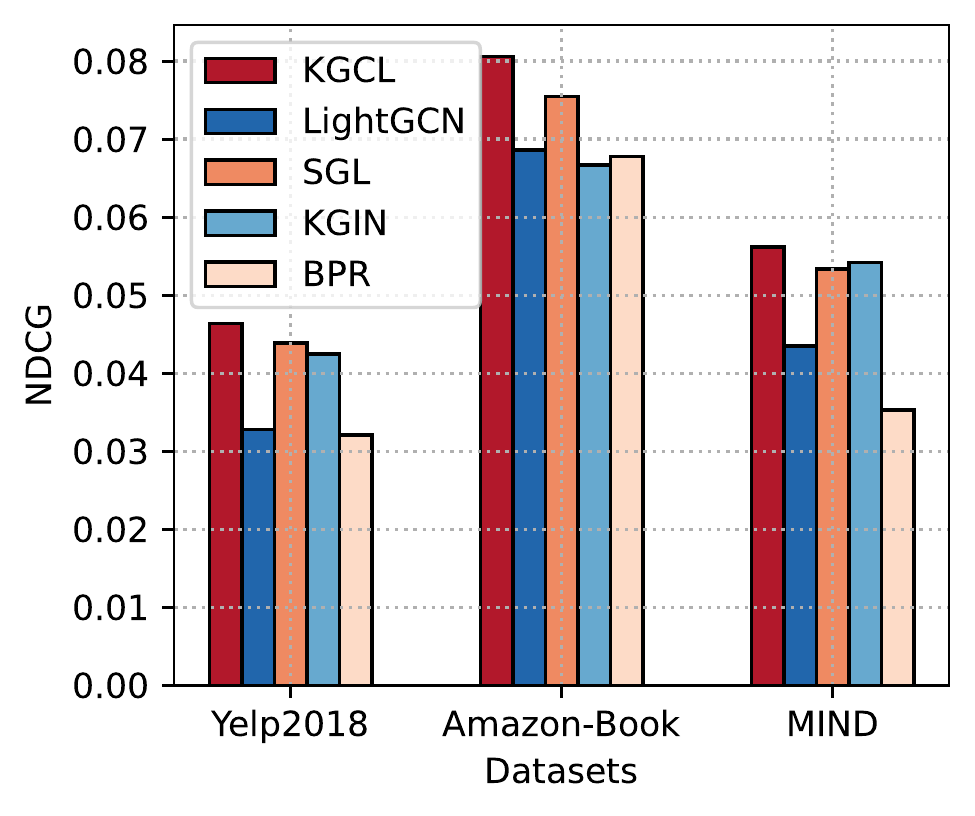}
	}
	\vspace{-0.2in}
	\caption{Comparison on cold-start users.}
	\label{figure:colduser}
	\vspace{-0.15in}
\end{figure}

\subsection{Benefits of \model\ in Alleviating Data Sparsity and Noise Effect (RQ3)}

In this subsection, we investigate the robustness of our \model\ by evaluating its performance for handling sparse and noisy data. \\\vspace{-0.12in}


\noindent \textbf{Sparse User Interactions}. To investigate the robustness of our \model\ in handing users without sufficient interactions, we follow similar settings in~\cite{yu2021self} to generate sparse user set with less than 20 interactions for Yelp2018 and Amazon-Book, and 5 interactions for MIND data. The results on sparse users are reported in Figure~\ref{figure:colduser}. \\\vspace{-0.1in}

\noindent \textbf{Long-tail Item Recommendation}. To justify the effect of our \model\ in long-tail item recommendation, we split all items into five groups with equal number of items (interaction density increases from group 0 to group 4). Separated evaluations are conducted on different item groups. Results are shown in Figure~\ref{figure:longtail}. \\\vspace{-0.12in}


\begin{figure}[ht]
	\subfigure[Recall]{
		\includegraphics[width=0.4\columnwidth]{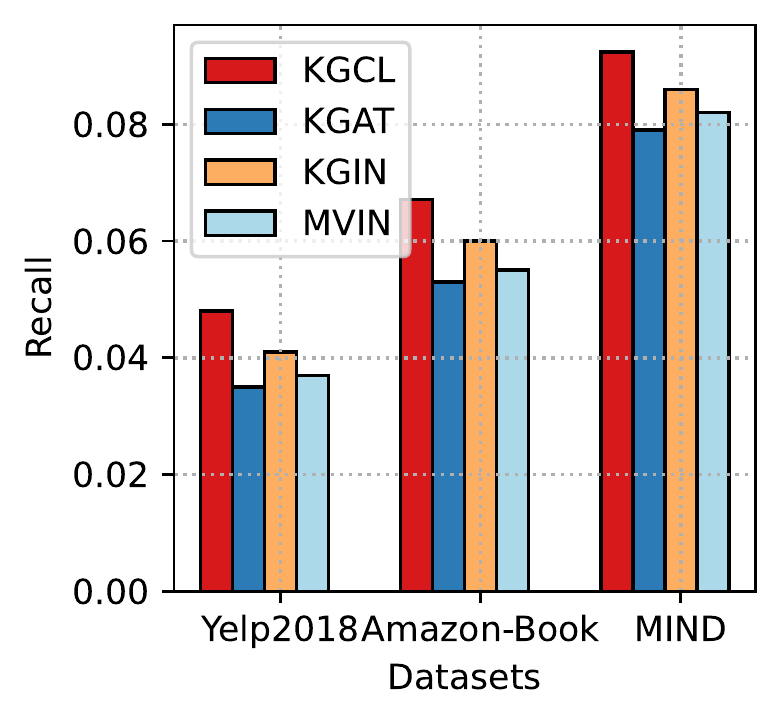}
	}
	\subfigure[NDCG]{
	    \includegraphics[width=0.4\columnwidth]{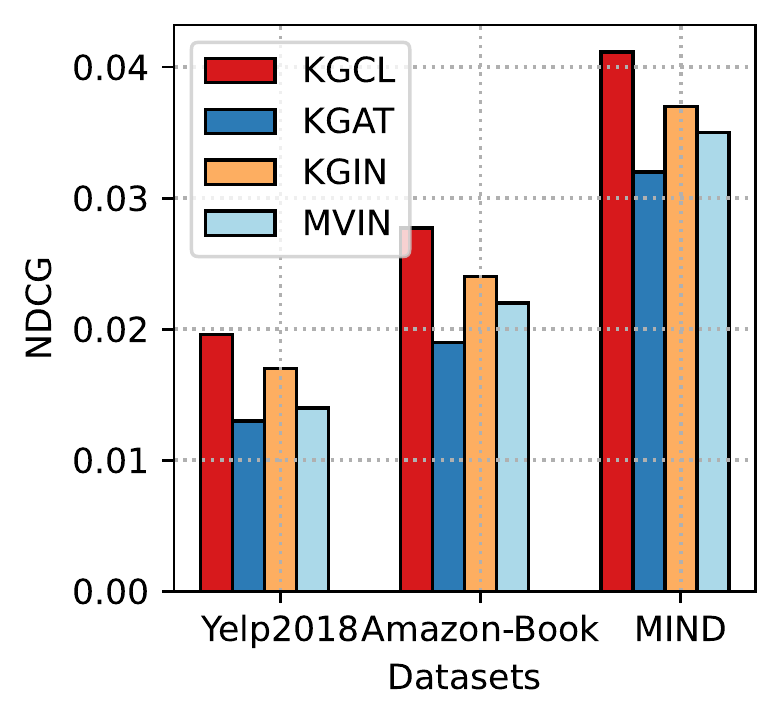}
	}
	\vspace{-0.2in}
	\caption{Recommendation performance comparison on items connected to long-tail KG entities.}
	\label{figure:coldkg}
	\vspace{-0.25in}
\end{figure}

\begin{table}[t]
	\centering
	\caption{Performance in alleviating KG noise.}
	\vspace{-0.15in}
	\label{results:adversarial}
	\resizebox{\linewidth}{!}{
	\begin{tabular}{c|cc|cc|cc|c}
		\hline
		\multirow{2}{*}{Model} & \multicolumn{2}{c|}{Yelp2018} & \multicolumn{2}{c|}{Amazon-book} & \multicolumn{2}{c|}{MIND} 
		& \multirow{2}{*}{\textbf{Avg. Dec.}}
		\\
		~ & Recall & NDCG & Recall & NDCG & Recall & NDCG & ~\\
		\hline
		\hline
		KGAT & 6.51\% & 0.0409 & 13.29\% & 0.0639 & 8.73\% & 0.0370 & 13.57\% \\
		KGIN & 6.85\% & 0.0444 & 13.69\% & 0.0719 & 10.32\% & 0.0527 & 3.37\% \\
		MVIN & 6.65\% & 0.0416 & 13.28\% & 0.0703 & 9.31\% & 0.0424 & 8.81\% \\ 
		\hline
		\hline
		\textbf{\baby} & \textbf{7.52\%} & \textbf{0.0490} & \textbf{14.93\%} & \textbf{0.0787} & \textbf{10.69\%} & \textbf{0.0550} & \textbf{0.58\%} \\
		\hline
	\end{tabular}
	}
	\vspace{-0.2in}
\end{table}

\noindent \textbf{Knowledge Graph Noise}. We investigate the robustness of our \model\ by separately injecting noisy triplets to the knowledge graph and testing on items connected with long-tail entities in the KG. In particular, we firstly randomly add 10\% noisy triplets into the existing KG data with the unchanged test set, to simulate the situation where the collected KG has a large number of topic-irrelevant entities. In addition, to simulate the KG noise scenario caused by long-tail entities, we collect the 20\% long-tail entities and filter the items connected with these long-tail entities in the testing data to perform evaluations. The results are reported in Table~\ref{results:adversarial} and Fig \ref{figure:coldkg}.
\begin{itemize}[leftmargin=*]

\item \model\ consistently outperforms SGL in all cases, which justifies the superiority of our knowledge graph-guided contrastive learning as compared to the randomly dropout-based strategy. Furthermore, the performance improvement of our \model\ over KGIN, indicates the necessity of incorporating knowledge graph-guided self-supervised signals into knowledge-aware recommender systems, so as to address the issue of noisy knowledge graph in misleading the encoding of user preference for recommendation. \\\vspace{-0.12in}

	

\item Considering the recommendation scenario in which items often exhibit long-tail distribution, our \model\ significantly improves the recommendation performance for long-tail items. This observation again demonstrates the superiority of our \model\ method in alleviating popularity bias for recommendation. However, unpopular items are less likely to be recommended by other baselines. Moreover, the performance superiority of our \model\ compared with competitive KG-enhanced recommender systems (\eg, KGAT, CKAN), indicates that blindly incorporating knowledge graph information into collaborative filtering may involve item relation noise, and cannot effectively alleviate popularity bias.\\\vspace{-0.12in}

\item Our \model\ always achieves the best performance when competing with state-of-the-art knowledge-aware recommendation models, in distilling useful information from noisy knowledge graph to assist the modeling of user preference. Specifically, \model achieves lowest average performance decreasing in alleviating KG noise (Table \ref{results:adversarial}), and the best evaluation results on items with sparse knowledge entities (Figure \ref{figure:coldkg}). This verifies the rationality of our knowledge graph contrastive learning paradigm in discovering relevant item semantics from noisy KG information.



\end{itemize}

\begin{figure}[t]
	\centering
	\includegraphics[width=0.95\columnwidth]{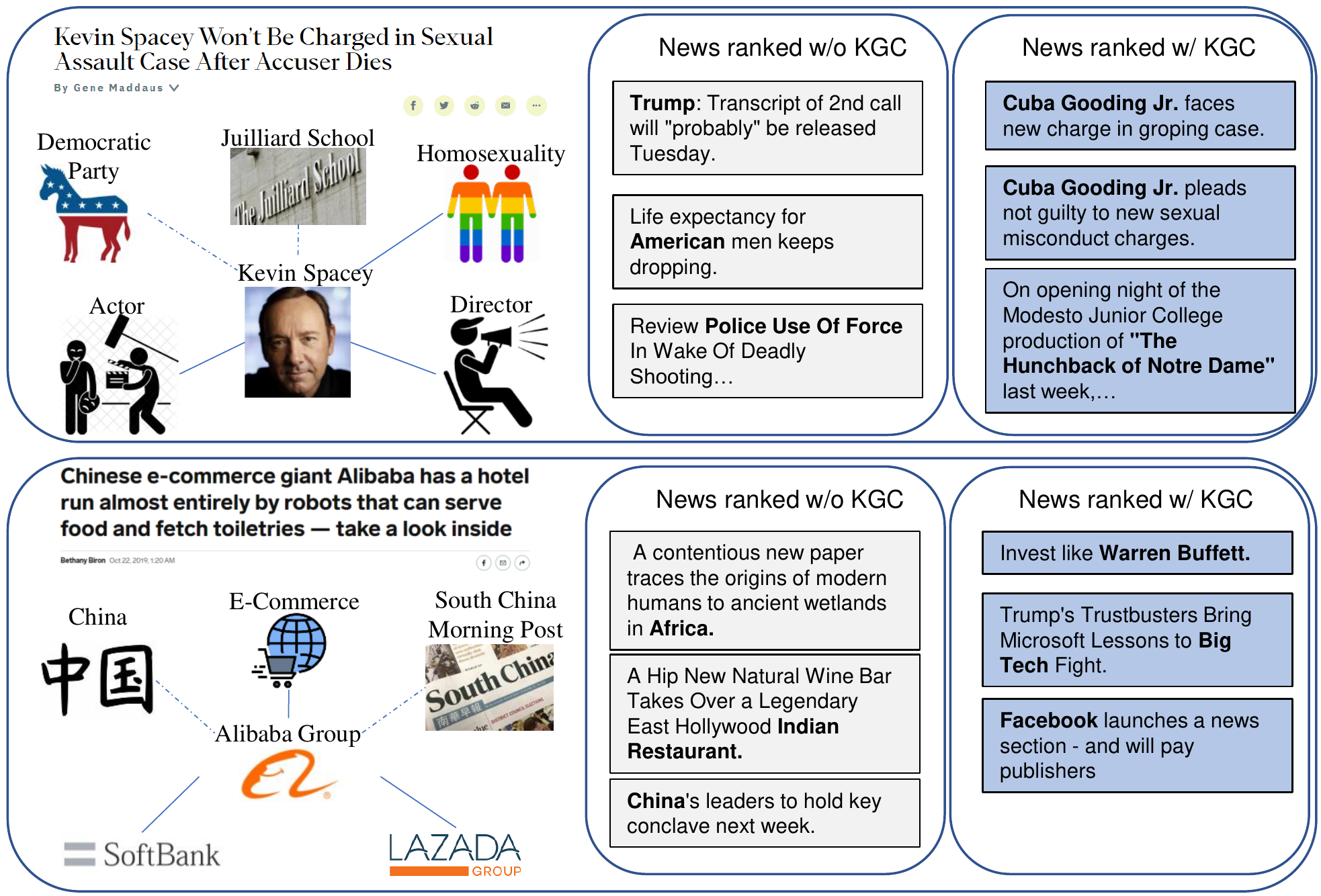}
	\vspace{-0.1in}
	\caption{Two examples of relevant news ranked w/ and w/o \model. News in blue color means semantically relevant and bold font means the entities extracted from the news item.}
	\label{figure:case}
	\vspace{-0.2in}
\end{figure}

\subsection{Case Studies (RQ4)}
\label{exp:case}


We perform case studies with sampled examples for news recommendation, to show the inference results with and without our knowledge graph contrastive learning (as illustrated in Figure \ref{figure:case}).


The first news case is about a celebrity \textit{Kevin Spacey} and his controversial affairs. We show entities connected with this news in the provided KG information, where his political position \textit{Democratic Party} and his graduated school \textit{Julliard School} are obviously irrelevant to this news. These noisy KG information may mislead the user representation by introducing biased information about politics or education. On the right part, we show three most similar news with this sampled news, ranked by the model with and without the KG-aware contrastive learning, respectively. Specifically, we can see that similar news ranked by model without KGC all relate to the theme of nationality and politics. Among them, the news topics are about \textit{Donald Trump}, \textit{American People} and \textit{Police Use of Force}, which are all irrelevant to the movie celebrity \textit{Kevin Spacey}. In contrast, the results inferred by our \model\ are all closely relevant to this news, \ie movie celebrity \textit{Cuba Gooding Jr.} or the movie \textit{The Hunchback of Notre Dame}. Coincidentally, \textit{Cuba Gooding Jr.} faces similar legal charges with \textit{Kevin Spacey}. After effectively denoising KG information, our method can correlate \textit{Cuba Gooding Jr.} and \textit{Kevin Spacey} with each other for accurate recommendation.


Another news example about China's tech giant \textit{Alibaba} is shown in the below part. Analogously, observed noisy entities \textit{China} and \textit{South China Morning Post} may influence the learning process of item-wise semantic relatedness towards the media press and nationality news. We can observe that the recommended similar news are all about national news from \textit{China}, \textit{India} and \textit{Africa}. By integrating our knowledge graph contrastive learning component, our model allows the recommendation framework to capture accurate semantic dependency among items by debiasing noisy entity-dependent relationships. In particular, the news identified by our \model\ is specific to the big tech companies, which are very relevant to the target news. Overall, \model\ is able to deconfound recommender system for alleviating knowledge graph information bias and eliminating the impact of irrelevant entities.

\section{Related Work}


\noindent \textbf{Knowledge Graph-enhanced Recommendation}.
Prior methods of KG-enhanced recommendation methods can be roughly grouped into two categories: embedding-based methods and path-based methods. For embedding-based methods \cite{zhang2016collaborative, wang2018dkn, xin2019relational, tian2021joint}, they leverage relations and entities in the KG to enhance the semantic representations in recommender systems. Usually, these methods apply a transition constrain to learn meaningful knowledge embedding for users and items. For example, CKE \cite{zhang2016collaborative} incorporates different types of side information into the collaborative filtering framework. In CKE model, the embedding of items' structural knowledge is encoded with TransR~\cite{lin2015learning}, and the textual and visual knowledge are learned with the proposed auto-encoder. Another representative method is DKN~\cite{wang2018dkn}, which integrates the semantic representations of news to learn better item embeddings.

\indent Path-based methods \cite{wang2019kgat,wang2019kgcn,wang2018ripplenet,xia2021knowledge, hu2018leveraging} aim to explore the potential information between items in knowledge graphs by constructing meta-path for information propagation. For instance, MCRec \cite{hu2018leveraging} designs meta-path-based mutual attention mechanism for Top-N recommendation, which produces user, item, and meta-path-based context representations. Overall, they offer relatively superior performance as compared to most embedding-based methods, since high-order knowledge-aware dependencies can be captured in those approaches. However, path-based methods highly depend on the design of meta-paths, which relies on the domain knowledge and human efforts. Moreover, aggregating information along different meta-paths is very time-consuming, resulting in inefficient knowledge-aware recommender systems. \\\vspace{-0.1in}


\noindent \textbf{Contrastive Learning for Recommender System.}
Recently, contrastive learning has attracted much attention in offering self-supervised signals for various domains, nature language processing~\cite{fu2021lrc} and image data analysis~\cite{deng2020disentangled}. It aims to learn quality discriminative representations by contrasting positive and negative samples from different views. Several recent attempts have brought the self-supervised learning to the recommendation \cite{wu2021self,liu2021contrastive,wei2022contrastive,long2021social}. For example, SGL~\cite{wu2021self} performs dropout operations over the graph connection structures with different strategies, \ie, node dropout, edge dropout and random walk. Additionally, CML~\cite{wei2022contrastive} enhances the recommender system with the consideration of multi-behavior relationships between users and items with contrastive learning. Motivated by these existing contrastive learning frameworks, this work develops a new graph contrastive learning paradigm for recommendation by effectively integrating knowledge graph representation and user-item interaction augmentation.
\section{Conclusion}

In this work, our proposed \model\ framework performs the initial attempts to explore the knowledge graph semantics and alleviate the data noise issue for recommendation under a knowledge-guided contrastive learning paradigm. The KG-aware data augmentation is conducted to investigate auxiliary self-supervised signals, based on estimating the effect of knowledge ambiguous items for user preference learning. This work opens up new research possibilities for knowledge-aware recommender systems. Extensive experiments on several real-world datasets have demonstrated the superiority of \model\ as compared to various state-of-the-art methods.



\section*{Acknowledgments}
This research is supported by the research grants from the Department of Computer Science \& Musketeers Foundation Institute of Data Science at the University of Hong Kong.

\clearpage

\bibliographystyle{ACM-Reference-Format}
\balance
\bibliography{reference}

\end{document}